\documentclass[useAMS,usenatbib]{mn2e}  
\usepackage{graphicx,epsfig,psfig}  
\usepackage{booktabs} 
\usepackage{kellymacros2} 
\usepackage{amsmath}
\usepackage[usenames]{color}

\long\def\Ignore#1{\relax}

\title[KIC\,3749404]{KIC 3749404: A \HB\ Star with Rapid Apsidal Advance Indicative of a Tertiary Component}

\author[Hambleton et~al.]{K. Hambleton$^{1,2}$\thanks{Email: kelly.hambleton$@$villanova.edu}, D. W. Kurtz$^{2}$, A. Pr\v sa$^{2}$, S. N. Quinn$^{3}$, J. Fuller$^{4}$, S. J. Murphy$^{5}$,\newauthor S. E. Thompson$^{6,7}$, D. W. Latham$^{8}$, A. Shporer$^{9}$\thanks{Sagan Fellow}\\
$^{1}$Department of Astronomy and Astrophysics, Villanova University, 800 East Lancaster Avenue, Villanova, PA 19085, USA\\
$^{2}$Jeremiah Horrocks Institute, University of Central Lancashire, Preston, PR1~2HE, UK\\
$^{3}$Department of Physics \& Astronomy, Georgia State University, 25 Park Place Suite 605, Atlanta, GA 30302, USA\\
$^{4}$TAPIR, Walter Burke Institute for Theoretical Physics, Mailcode 350-17\\
$^{5}$Stellar Astrophysics Centre, Department of Physics and Astronomy, Aarhus University, DK-8000 Aarhus C, Denmark\\
$^{6}$NASA Ames Research Center, Moffett Field, CA 94035, USA\\ 
$^{7}$SETI Institute, 189 Bernardo Avenue Suite 100, Mountain View, CA 94043, USA\\
California Institute of Technology, Pasadena, CA 91125, USA\\
$^{8}$Harvard-Smithsonian Center for Astrophysics, 60 Garden St, Cambridge, MA 02138, USA\\
$^{9}$Jet Propulsion Laboratory, California Institute of Technology, 4800 Oak Grove Drive, Pasadena, CA 91109, USA\\
}

\begin{document}
\date{Accepted} 
\pagerange{\pageref{firstpage}--\pageref{lastpage}} \pubyear{2016} 

\maketitle 
\label{firstpage}

\begin{abstract}
Heartbeat stars are eccentric ($e > 0.2$) ellipsoidal variables whose light curves resemble a cardiogram. We present the observations and corresponding model of KIC\,3749404, a highly eccentric ($e$\,=\,0.66), short period ($P$\,=\,20.3\,d) \hb\ star with tidally induced pulsations. A binary star model was created using \ph, which we modified to include tidally induced pulsations and Doppler boosting. The morphology of the photometric periastron variation (\hb) depends strongly on the eccentricity, inclination and argument of periastron. We show that the inclusion of tidally induced pulsations in the model significantly changes the parameter values, specifically the inclination and those parameters dependent on it. Furthermore, we determine the rate of apsidal advance by modelling the periastron variation at the beginning and end of the 4-yr \kep\ data set and dividing by the elapsed time. We compare the model with the theoretical expectations for classical and general relativistic apsidal motion and find the observed rate to be two orders of magnitude greater than the theoretical rate. We find that the observed rate cannot be explained by tidally induced pulsations alone and consequently hypothesise the presence of a third body in the system. 

\end{abstract}


\section{Introduction} 
\label{sec:intro}

\HB\ stars are an interesting class of eccentric ellipsoidal variables introduced by \citet{Thompson2012}. The study of \hb\ stars was initiated with the discovery of KIC\,8112039 (also known as KOI-54, where KOI stands for \kep\ Object of Interest; \citealt{Welsh2011}) and subsequent theoretical papers on this iconic object \citep{Fuller2012,Burkart2012,Oleary2014}. The most prominent feature in the light curve of KOI-54, and all \hb\ stars, is the variation in brightness at periastron, which is a consequence of stellar deformation caused by tides and mutual irradiation. The morphology of this feature primarily depends on the argument of periastron, eccentricity and inclination of the object, as described by \citet{Kumar1995}. For most \hb\ stars, irradiation is a second-order effect in the light curve; however, for KOI-54 and other objects with components of similar size and with a small periastron distance, irradiation can also be dominant in the light curve (e.g., it is about 50 per cent of the \hb\ amplitude of KOI-54). The amplitude of the periastron variation also depends on the periastron distance, mutual irradiation and effects such as gravity darkening, making detailed models necessary.

With the advent of highly precise observations from satellites such as {\it Kepler} \citep{Borucki2010, Gilliland2010, Batalha2010}, {\it MOST} \citep{Walker2003} and {\it CoRoT} \citep{Baglin2006}, it is obvious that these objects are not as rare as previously thought. To date, with the help of Planet Hunters, the \kep\ Eclipsing Binary Working Group and the \kep\ Science Office, we (the Heartbeat stars team) have identified \hbno\ \hb\ stars in the \kep\ data. An up-to-date list of \kep\ \hb\ stars can be found at the Kepler Eclipsing Binary web page\footnote{\sf http://keplerebs.villanova.edu}\citep{Kirk2015}. \HB\ stars have also been observed using other projects, including two by the CoRoT mission \citep{Maceroni2009,Hareter2014}, and eight by the ground-based Optical Gravitational Lensing Experiment, OGLE \citep{Nicholls2012}. For details of \hb\ stars observed using the \kep\ satellite, see \citet{Welsh2011, Thompson2012, Hambleton2013, Beck2014, Schmid2015, Smullen2015, Shporer2016}.

\begin{figure*}
\hfill{}
\includegraphics[width=\hsize, height=10cm]{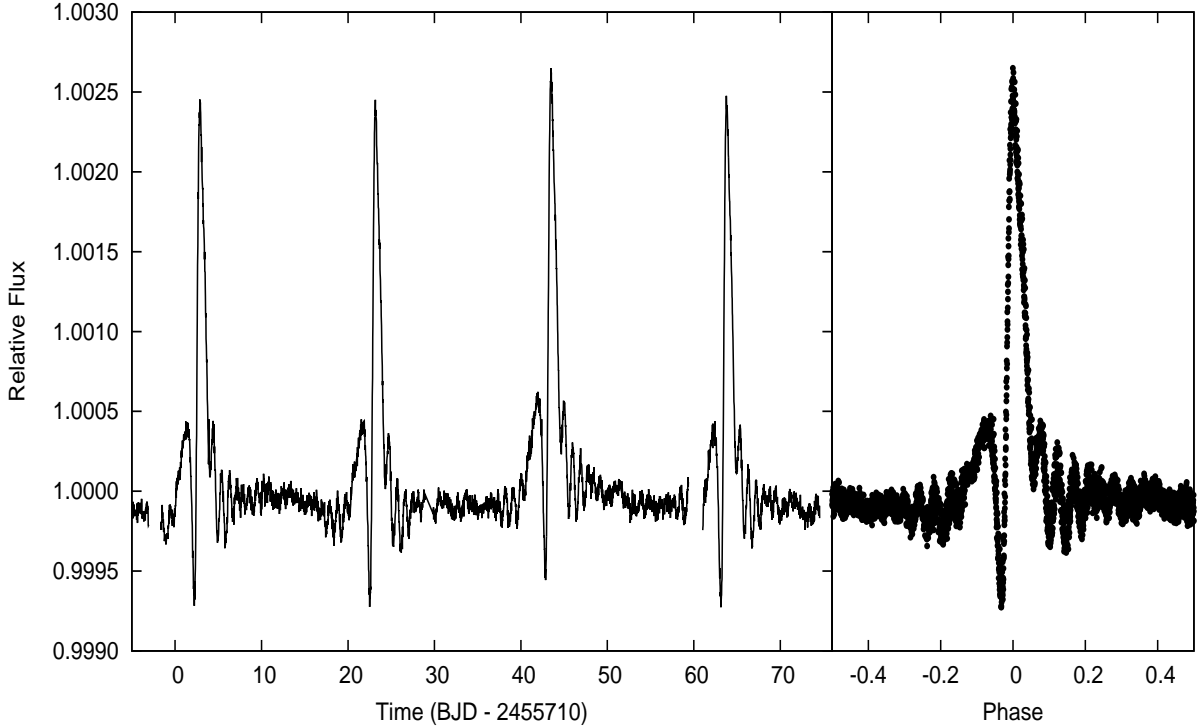}\\
\small\caption{{\it Kepler} long-cadence (Quarters 9--10) time series (left) and the phase folded light curve (right) of KIC\,3749404. The dominant feature of the light curve, repeating once per orbital cycle (20.3\,d), is the variability due to tidal deformation. Tidally excited modes, the smaller variations, are also clearly visible in the phase folded light curve due to their commensurability with the orbital period.}
\label{fig:lc}
\hfill{}
\end{figure*}

Due to the strong and variable gravitational interactions between stellar components, a subset of these objects exhibit tidally excited modes. These occur when the forcing frequency of the tide is close to a stellar eigenfrequency, significantly increasing the amplitude of the mode. Caused by resonances with dynamical tides, these modes were hypothesised by \citet{Cowling1941} and \citet{Zahn1975} to be the mechanism for orbital circularisation. Prior to the launch of \kep, tidally excited modes had only been identified in a handful of objects (e.g., HD177863; \citealt{Willems2002}). However, in the \kep\ data alone, we have identified 24 objects showing obvious, high amplitude tidally induced pulsations, and following closer inspection estimate that $\sim$20 per cent of our sample pulsate with tidally excited modes.

KIC\,3749404 is a binary star system containing an A and an F star in a close ($P$\,=\,20.3\,d), highly eccentric (e\,$\sim$\,0.66) orbit. The work of \citet{Smullen2015} contains a catalog of \hb\ star radial velocities and their corresponding models for 7 objects including KIC\,3749404. We expand on this work by providing a complete model of the \kep\ light curve and radial velocity data. A list of identifiers and basic data for KIC\,3749404 can be found in Table\,\ref{tab:ID}. KIC\,3749404 was selected for detailed study due to its interesting light curve morphology and prominent tidally excited pulsations, which can be seen in Fig.\,\ref{fig:lc}. The signature of tidally excited modes are frequencies that are precise integer multiples of the orbital frequency. This feature is obvious in KIC\,3749404, as the phased data clearly show the pulsations -- this only occurs when the number of pulsations per orbit is very close to an integer. 

\begin{table}
\caption{
\label{tab:ID}
Identifiers and basic data for KIC\,3749404. The \kep\ magnitude specified is derived from the \kep\ broadband filter.} 
\begin{center}
\begin{tabular}{l l}
\hline
\multicolumn {2}{l}{Identifiers}\\\hline
KIC                    & \ \ 3749404\\
TYC                    & \ \ 3134-165-1 \\
GSC                    & \ \ 03134-00165\\
2MASS                  & \ \ J19281908+3850135\\\hline
\multicolumn {2}{l}{Position and magnitudes}\\\hline
RA \ (J2000)           & \ \ 19:28:19.0894\\
Dec (J2000)            & +38:50:13.603 \\
{\it V}                & \ \ 10.6 \\
{\it B}                & \ \ 10.9 \\
Kp                     & \ \ 10.6\\
\hline
\end{tabular}
\end{center}
\end{table}

Apsidal motion is the rotation of the line of apsides about the binary centre of mass, which, in \hb\ stars, is seen as a change in the shape of the ellipsoidal variation (\hb) over time. Classical apsidal motion occurs when gravitational interactions generate a tidal bulge on the surface of a star or stellar rotation causes a star to become oblate. This deviation from stellar point masses causes the orbit to precesses about the center of mass. Apsidal motion is additionally caused by General Relativity. KIC\,3749404 demonstrates rapid apsidal motion. This is an interesting feature, as a discrepancy has been found in several objects between the central density parameter ($k_2$) determined through apsidal advance and that predicted by models, e.g. see \citet{Gimenez1982}. While many of these cases can be attributed to the inadequate treatment of stellar rotation, or imprecise stellar radii (as the theoretical rate of classical apsidal advance scales with $R^{-5}$), as shown by \citet{Claret1993a}, there are some cases where theory and observation do not agree, e.g. \citet{Guinan1985}. One reason for this discrepancy could be the presence of tidally induced pulsations. \citet{Papaloizou1980} theorised that tidally induced pulsations could alter the rate of apsidal advance, and \citet{Claret2003} further showed that the degree of the discrepancy is likely associated with the phase of the resonance relative to the orbit. For KIC\,3749404, we find that the level of discrepancy is, however, unlikely to be a consequence of tidally induced pulsations alone. We thus hypothesise that the rapid apsidal motion is a consequence of a tertiray component in the system.

\section{Observations}
\label{sec:Obs}

\subsection{{\it Kepler\/} Photometry}
\label{sec:kep_phot}

The observations of KIC\,3749404 consist of both long cadence (LC) data, during Quarters 0--17, and short cadence (SC) data, during Quarters 3.3 and 11--17. {\it Kepler} Quarters are variable in time span, but typically are about 93\,d, or one quarter of a complete 372.5-d orbit around the Sun \citep{Kjeldsen2010}. LC data correspond to a sampling rate of 29.4244\,min and SC data to a sampling rate of 58.8488\,s. The \kep\ photometric observations that have been analysed for KIC\,3749404 span from 2009 May -- 2013 May. All observations were obtained from the Mikulski Archive for Space Telescopes and were a part of Data Releases 21--23 \citep{DRN21, DRN22, DRN23}. We used the long cadence data to obtain a model of the binary features and the g-mode frequencies, and to determine the rate of apsidal advance. We used the SC data of Quarters\,11--17 to look for the presence of pressure modes (p~modes). We identified possible signatures of low-amplitude unresolved p~modes at $\sim$31\,d$^{-1}$. However, as these are in the region of 4 known artifacts, 31.00\,d$^{-1}$, 31.10\,d$^{-1}$, 31.35\,d$^{-1}$, 31.61\,d$^{-1}$ \citep{DCH2013} and because we do not see them reflected about the Nyquist frequency in the long-cadence data, we conclude that there are no p~modes in the short cadence light curve of KIC\,3749404. 

The photometric observations were made using the \kep\ broadband filter, which is essentially a white light filter. When selecting the type of \kep\ product to use we noted that the msMAP version of the PDC (pre-search data conditioning) \kep\ pipeline does not preserve periodicities greater than $\sim$20\,d \citep{DCH2013}. As the orbital period of KIC\,3749404 is $\sim$20\,d, we elected to use the simple aperture photometry data (instead of PDC) to ensure that no information had been removed by the \kep\ pipeline, which is fine-tuned for transiting planet detection. 

As each \kep\ pixel is 4\,$\times$\,4\,arcsec, it is possible that some contamination will occur within the photometric field. The contamination value for KIC\,3749404, specified by the Kepler Asteroseismic Science Operations Centre (KASOC), is estimated to be 0.011, where 0 implies no contamination and 1 implies complete contamination of the CCD pixels. This contamination value suggests that KIC\,3749404 suffers minimally from third light, if at all. We applied the {\it pyKE} tools \citep{Still2012} to the target pixel files to assess the flux incident on each individual pixel. From this we determined that the contamination level for KIC\,3749404 is negligible.

As \hb\ stars are notoriously difficult to detrend, we tried several different methods of detrending for this object including polyfits of different lengths and orders, and cotrending basis vectors. The data were detrended using third order polynomials that were applied between breaks in the data, using the {\it kephem} software, as this method yeilded the best results. {\it Kephem} \citep{Prsa2011}, an interactive graphical user interface package that enables the deterending of data using Legendre polynomials and further incorporates 3 methods of period analysis: Lomb-Scargle (LS; Lomb 1976; Scargle 1982),\nocite{Lomb1976, Scargle1982} Analysis of Variance (AoV; Schwarzenberg-Czerny 1989),\nocite{Schwarzenberg-Czerny1989} and Box-fitting Least Squares (BLS; Kov{\'a}cs et al. 2002),\nocite{Kovacs2002} as implemented in the {\it vartools} package (Hartman et al. 2008)\nocite{Hartmann1998}. 

We further cleaned the data by removing obvious outliers. For the determination of the binary and pulsation parameters we used Quarters 8--10. We did not use the total data set as the periastron variation is changing in time due to apsidal motion, which causes smearing of the periastron variation. We chose to use Quarters\,8--10 as these data were observed simultaneously with the spectral observations and three Quarters provide an acceptable balance between the number of orbits and the smearing of the periastron variation in the phase folded light curve due to apsidal motion. 

Due to computational costs, the total number of data points was then reduced from 12\,361 to 1\,436. To effect the reduction, we assigned each data point with a random number from 0 to 1, and removed all points above a specified threshold: 0.25 for the periastron variation and 0.12 for all other points. To avoid having discrete jumps in the number of data points at the transition regions, we applied sigmoid functions so that the number of data points was gradually increased/decreased. We elected to use random selection over binning the data to avoid weighting the data towards spurious trends. We determined the per-point uncertainty to be $\sigma = 8.2$\,ppm by finding the standard deviation of segments of the data (away from the periastron variation) with the initial model and harmonics of the orbital frequency removed. This was repeated for 10 segments and the results averaged. 

\begin{table}
\caption{
\label{tab:rvs}
Radial velocities of the two components of KIC\,3749404, determined using \tdcr. The spectra were observed using the Tillinghast Reflector Echelle Spectrograph on the 1.5-m telescope at the Fred L.\ Whipple Observatory between UT 2011 May 10 and 2011 Jun 23. The uncertainties on the primary and secondary radial velocities are $\sigma = 1.0$\,$\kms$ for all measurements. }
\begin{center}
\begin{tabular}{c r r} 
\hline
\multicolumn{1}{c}{BJD} & \multicolumn{1}{c}{Primary}  & \multicolumn{1}{c}{Secondary}  \\
\multicolumn{1}{c}{2450000.0 +}&\multicolumn{1}{c}{$\kms$} & \multicolumn{1}{c}{$\kms$}  \\
\hline
%&$\pm 1.0$ & $\pm 1.0$  \\
5691.9395&  -41.0&   19.8    \\
5692.9411&  -89.3&   84.8    \\
5693.9409&  -71.0&   60.3    \\
5694.8716&  -55.2&   38.0    \\
5695.9629&  -42.6&   21.0    \\
5696.8644&  -34.1&   10.6    \\
5697.8724&  -26.6&    1.2    \\
5698.9011&  -19.9&  -10.1   \\
5699.7996&  -15.7&  -15.0   \\
5701.8281&   -6.6&  -26.0    \\
5704.8259&    4.9&  -42.6    \\
5705.8248&    8.6&  -47.9    \\
5727.8552&   14.2&  -55.0   \\
5728.9439&   17.6&  -58.6   \\
5729.8041&   18.8&  -62.1   \\
5731.9046&    0.3&  -36.6    \\
5732.8222&  -64.9&   50.5   \\
5733.7524&  -86.9&   80.2   \\
5734.9330&  -63.9&   49.8   \\
5735.7603&  -52.0&   33.1   \\
\hline
\end{tabular}
\end{center}
\end{table}

\subsection{Binary Orbital Period Determination}

Using {\it kephem} (as described in \S\ref{sec:kep_phot}), period analysis was performed on all the LC data (Quarters 0--17) to determine the period of the binary orbit and BJD$_0$, the zero point in time (the maximum of the periastron variation). Due to the apsidal motion in the system, it is important to note that the period is the anomalous period and not the sidereal period. The ephemeris was found to be:\newline
\newline
Min\Rmnum{1} = BJD 2455611.342(3)+20.30635(8) $\times$ E\newline
\newline
\noindent
where Min\Rmnum{1} is the orbital ephemeris and the values in the parentheses give the one sigma uncertainties in the last digit. The period uncertainty was obtained by applying an adaptation of the Period Error Calculator algorithm of \citet{Mighell2013}, as specified in \citet{Kirk2015}. 

\subsection{Ground-based Spectroscopy}
\label{sec:spec}

We followed-up KIC\,3749404 spectroscopically to measure the radial velocities of the two components and to characterize the stellar
atmospheres. A total of $20$\ observations were collected with the Tillinghast Reflector Echelle Spectrograph \citep[TRES;][]{Furesz2008}
on the 1.5-m telescope at the Fred L.\ Whipple Observatory between UT 2011 May 10 and 2011 June 23. The spectra cover the wavelength 
between $\sim 3900$\ and $9000$\,\AA\ at a resolving power of $R \approx 44\,000$. We extracted the spectra following the procedures
outlined by \citet{Buchhave2010}.

We derived the radial velocities for both stars from the TRES spectra using the
two-dimensional cross-correlation technique \tdcr\ \citep{Zucker1994},
with synthetic spectral templates generated from Kurucz model
atmospheres. \tdcr\ uses two dimensional cross-correlation to identify the optimal radial velocities of the two input model spectra with respect to the observed data. We used one echelle order of about $100$\,\AA\ centered on
$5190$\,\AA, which includes the gravity-sensitive Mg\,{\Rmnum 1}\,b
triplet. The radial velocities are reported in Table\,\ref{tab:rvs}. For discussion on the determination of the spectroscopic parameters and the case for the Am nature of the primary star, see \S\ref{sec:am}.

\section{Binary Modelling}\label{sec:mod}

\subsection{Tidally Excited Modes\label{sec:tidal}}

Stellar pulsation modes may be tidally excited when a multiple of the binary orbital frequency is close to a stellar eigenfrequency. The signature of a tidally excited mode is a pulsation frequency that is a precise multiple of the orbital frequency. In the initial analysis of KIC\,3749404 we identified 7 pulsations that are multiples of the orbital frequency with amplitudes greater than 20\mumag. The nature of tidally excited pulsations makes them intrinsically difficult to extract: their frequencies are multiples of the orbital frequency and hence they overlap with orbital harmonics created by the Fourier decomposition of the near-periastron heartbeat signal. This is especially problematic for relatively low-frequency tidally excited modes, which are common. For this reason we added the capability to \ph\ to model the binary features and pulsations simultaneously by combining the binary star model with sine waves at frequencies that are multiples of the orbital frequency. To our knowledge, this is the first time that pulsations have been modelled simultaneously with light curve and radial velocity data. We further incorporated Markov chain Monte Carlo methods to fit our model to the data. 

Prior to fitting the binary star features and pulsation parameters simultaneously, we identified the prominent pulsations in the light curve. To do this, we generated the residuals to the initial binary fit; masked the region of the periastron variation (between phases -0.19--0.12) to reduce the impact of an imperfect model/over-fitting the pulsations; and applied a Fourier transform to the residual data. 

\begin{figure*}
\hfill{}
\includegraphics[width=\hsize, height=10cm]{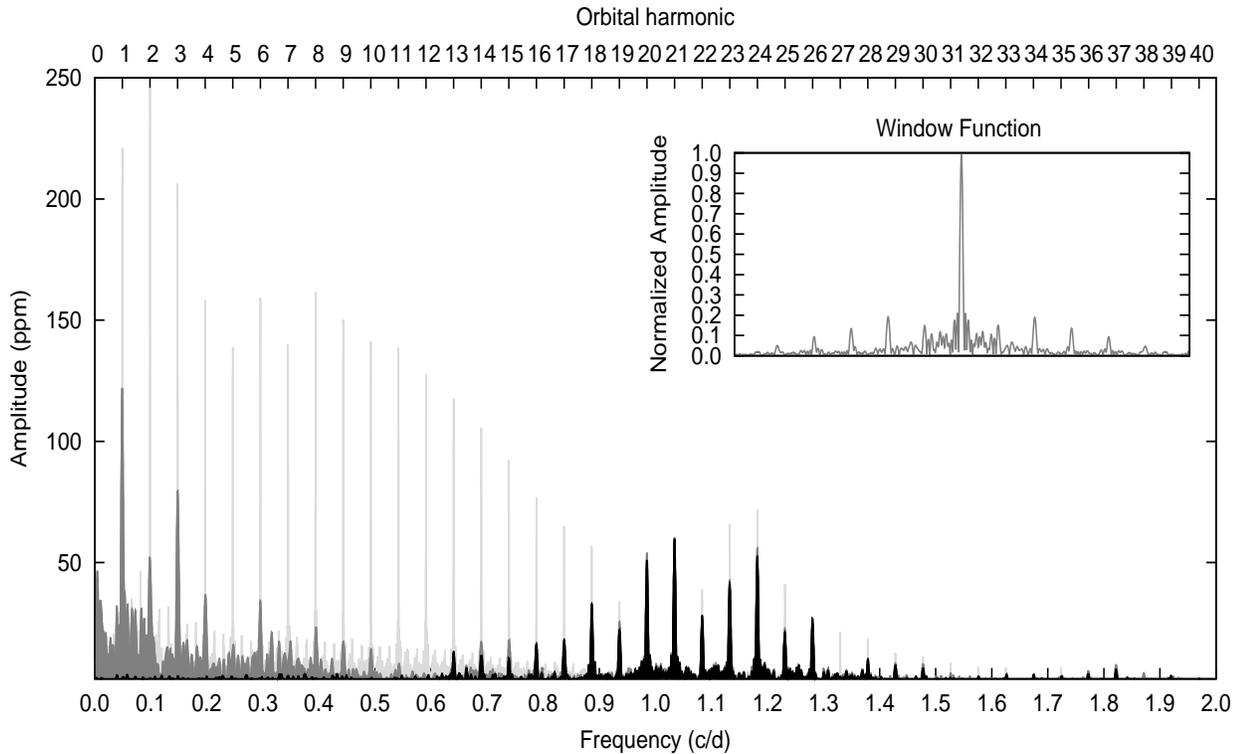}\\
\small\caption{A Fourier transform of the original detrended data (light grey); data with the initial binary star model removed (grey); and with the initial binary star model and frequencies less than than 0.5\perday (down to an amplitude of 4\mumag) removed (black). The latter, black Fourier transform contains tidally induced pulsations only. The insert depicts the window pattern of the highest amplitude peak (21\nuorb). All the peaks are at precise multiples of the orbital frequency.}
\label{fig:FT}
\hfill{}
\end{figure*}

The Fourier transform of the original, detrended data contains a double peaked distribution of frequencies where all the frequencies are multiples of the orbital frequency (cf. Fig.\,\ref{fig:FT}). The first peaked distribution of frequencies describe the binary star features in the light curve, including the periastron variation, and the second is caused by tidally induced pulsations. To further analyse the pulsations, we removed an initial binary model from the data and prewhitened any remaining frequencies below 0.5\perday, most of which are instrumental in nature, to an amplitude of 4\mumag. 
From the Fourier transform of the residual data, with the low frequency peaks removed, we identified all remaining significant harmonics from the data (down to 20\mumag) and fitted them simultaneously using linear least-squares. In order of amplitude, the identified frequencies are: (21, 24, 20, 23, 22, 25, 18) $\times$\,$\nu$$_{orb}$. For a more detailed discussion of the tidally induced pulsations, including the final pulsation parameters from the binary star model, see \S\ref{sec:tidal}.

The majority of \hb\ stars, including KIC\,3749404, do not have eclipses. We see a dip in the light curve of KIC\,3749404 because at that phase the stars are tidally distorted due to gravitational affects and we are viewing both stars with a smaller surface area relative to their surface area during the rest of the orbit. As there are no eclipses present in the light curve of KIC\,3749404, we are unable to determine from which component the tidally induced pulsations originate. However, as they form a single peaked distribution in the Fourier transform, we conclude that they originate in one star, although which star is currently not known.

\subsection{Simultaneous Binary and Pulsation Modelling\label{sec:binary}}

We applied the binary modelling code {\sc phoebe} \citep{Prsa2005}, which is an extension of the Wilson-Devinney code \citep{Wilson1971,Wilson1979,Wilson2004}, to the light curve of KIC\,3749404. {\sc phoebe} combines the complete treatment of the Roche potential with the detailed treatment of surface and horizon effects such as limb darkening, reflection and gravity brightening to derive an accurate model of the binary parameters. The current implementation uses the Wilson-Devinney method of summing over the discrete rectangular surface elements, which cover the distorted stellar surfaces, to determine an accurate representation of the total observed flux and consequently a complete set of stellar and orbital parameters. {\sc phoebe} incorporates all the functionality of the Wilson-Devinney code, but also provides an intuitive graphical user interface alongside many other improvements, including updated filters and bindings that enable interfacing between \ph\ and {\sc python} (see \S\,\ref{sec:bayes}).

\begin{table}
\caption{Fixed parameters and coefficients for the {\sc phoebe} best-fit model to 
the {\it Kepler} light curve for Quarters\,8--10. The values in the parentheses specify the one sigma uncertainties in the previous digit. The effective temperatures were determined using the spectral analysis performed in \S\ref{sec:am}. The binning undertaken with this analysis did not preserve \vsini. We therefore used the \vsini\ measurements obtained from our \tdcr\ analysis in \S\ref{sec:spec}.\label{tab:ParamFix}
}
\begin{center}
\begin{tabular}{||l|r||}
\hline
Parameter & Values\\
\hline 
Primary T$_{\mathrm{eff}}$ (K)                          & 8000(300)\\
Secondary T$_{\mathrm{eff}}$ (K)           	  	& 6900(300)\\
Primary $v\sin i$ ($\kms$)                              & 29(2)\\ 
Secondary $v\sin i$ ($\kms$)       			& 9(2)\\ 
Orbital Period (d)              			& 20.30635(15)\\
Time of primary minimum (BJD) 				& 2455611.342(3)\\
Primary Bolometric albedo       			& 1.0\\ 
Secondary Bolometric albedo   				& 0.6\\
Third light                     			& 0.000(6)\\
\hline
\end{tabular}
\end{center}
\end{table}
 
We calculated the value of $F$ for each component, where $F$ is the ratio of the rotational to orbital period. This was done for each component at each iteration by combining the spectroscopically determined $v \sin i$ values ($v_1\sin i = 29 \pm 2$\,km~s$^{-1}$ and $v_2\sin i  = 9 \pm 2$\,km~s$^{-1}$) with the model-determined values of the inclination and radii. We further fixed the stellar albedos to $A_1$\,=\,1.0 and \,$A_2$\,=\,0.6, which are theoretically predicted for stars with radiative and convective outer envelopes, respectively \citep{Rucinski1969a,Rucinski1969b}. We elected to fix the period and zero point in time (time of photometric maximum) to the values determined using {\it kephem} (see Table\,\ref{tab:ParamFix}), as the Lomb-Scargle method is more accurate than \ph\ for ephemeris determination. 

Following the work of \citet{Diaz-Cordoves1992} and \citet{vanHamme1993}, who showed that the square-root and logarithmic limb darkening laws are preferable for objects that radiate towards the IR and UV, respectively, we elected to use the logarithmic limb darkening law, as within \ph\ this is a system-wide parameter, and the primary star contributes the larger fraction of light to the system. 

\subsection{Posterior Determination of the Binary Star Parameters \label{sec:bayes}}

\begin{table}
\caption{Mean values of the fitted and calculated parameters determined using \ph\ with \emcee\ for 0 and 7 pulsations. The values are derived from Gaussian fits to the posterior distributions. The rotation rate refers to the number of rotations per orbit and the primary fractional luminosity is $L_1/(L_1+L_2)$. The values in parentheses give the uncertainty in the previous digit. 
\label{tab:ParamFree}
}
\begin{center} 
\begin{tabular}{l r r } 
\hline 
Parameter & \multicolumn {2}{c}{Number of pulsations}\\
          &  \multicolumn {1}{c}{0}& \multicolumn {1}{c}{7}\\ 
\hline 
\multicolumn {2}{l}{Fitted}\\
\hline
Orbital inclination (degrees)     & 60.2(1)  & 62(1)\\
Argument of periastron (rad)      & 2.14(2)  & 2.15(4)\\
Eccentricity                      & 0.633(3) & 0.659(6)\\
Primary gravity brightening       & 0.95(3)  & 0.95(3)\\
Secondary gravity brightening     & 0.29(2)  & 0.52(6)\\
Primary polar radius (\Rsun)      & 1.9(1)   & 1.98(4)\\
Secondary polar radius (\Rsun)    & 1.1(1)   & 1.20(4)\\
Mass ratio                        & 0.73(1)   & 0.739(9)\\
Phase shift                       & -0.110(3) & -0.109(5)\\
Gamma velocity (km~s$^{-1}$)      & -14(2)    & -15(1)\\
Semi-major axis (\Rsun)           & 43(1)     & 45.7(7)\\
\hline
\multicolumn {2}{l}{Calculated}\\
\hline
Primary mass ($\Msun$)            & 1.5(1)    & 1.78(6)\\
Secondary mass ($\Msun$)          & 1.1(1)    & 1.32(4)\\
Primary fractional luminosity     & 0.85(4)   & 0.82(2)\\
Primary rotation rate             & 7.1(3)    & 7.1(2)\\
Secondary rotation rate           & 4.0(1)    & 4.0(1)\\

\hline
\end{tabular} 
\end{center} 
\end{table}

\begin{figure*} 
\includegraphics[width=\hsize]{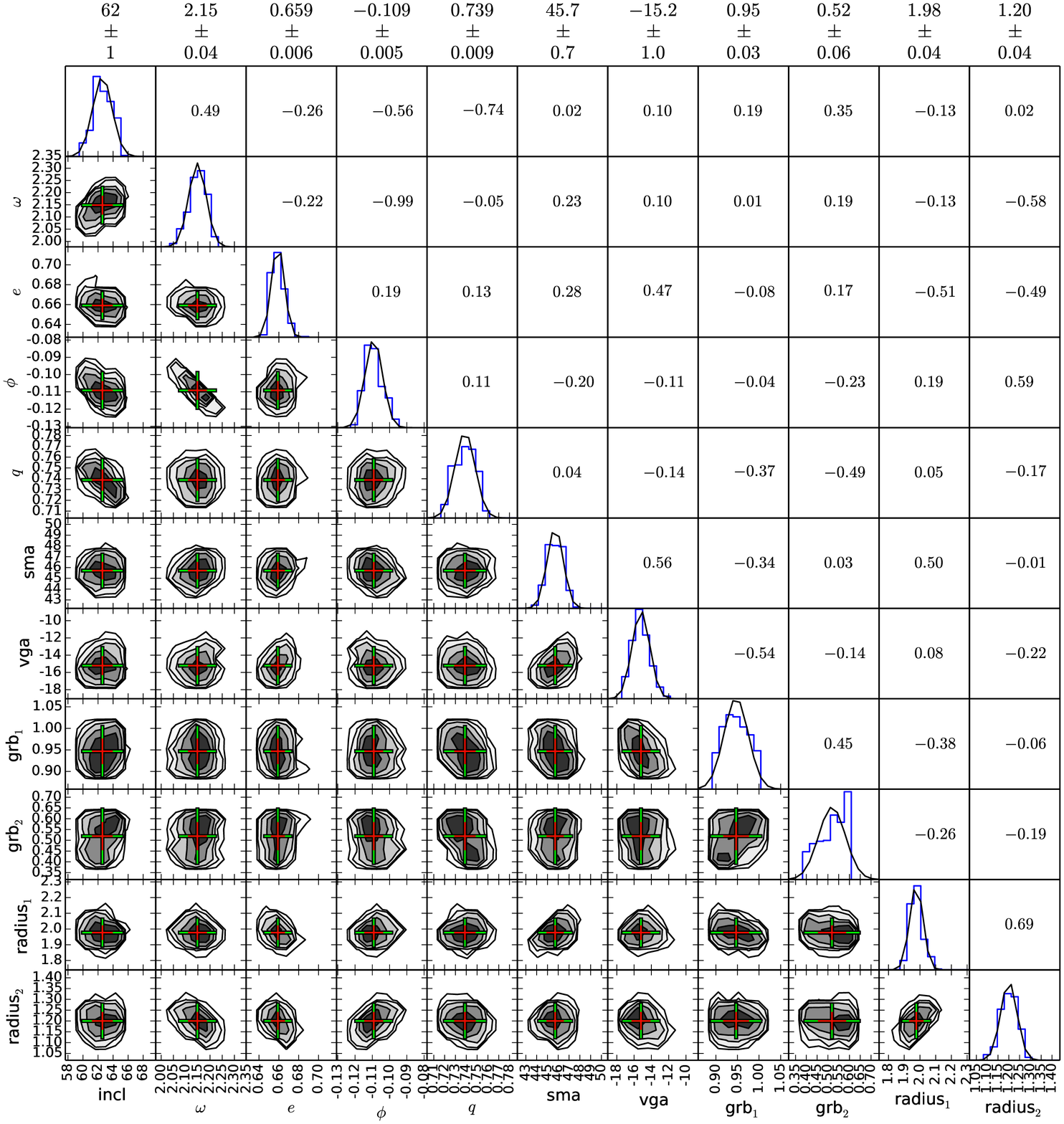}  
\small\caption{The posterior distributions of the binary star parameters for the model with seven pulsations, where $incl$ is the inclination of the binary star orbit in degrees; $\omega$ is the argument of periastron in radians; $e$ is the eccentricity; $\phi$ is the phase shift; $q$ is the mass ratio ($m_2/m_1$); sma is the semi-major axis in \Rsun; vga is the gamma velocity in $\kms$; grb$_1$ and grb$_2$ are the gravity brightening exponents of the primary and secondary components, respectively; and radius$_1$ and radius$_2$ are the primary and secondary polar radii in \Rsun, respectively. We elected to calculate the polar radii since they are more constant throughout the orbit (as opposed to the point radii). Lower left sub-plots: two dimensional cross sections of the posteriors. The crosses show the one sigma (red) and two sigma (green) uncertainties, and are centred on the minima. Diagonal sub-plots from top left to bottom right: histograms displaying the posterior distribution of each individual parameter. Upper right values: the correlations for the two-dimensional cross sections mirrored along the diagonal, where 1 is complete correlation, -1 is a complete anti-correlation and 0 is no correlation. The values above the plot give the mean value and one sigma uncertainty for each parameter, based on the fitted Gaussians.} 
\label{fig:posteriors} 
\end{figure*} 

To determine the posteriors of the binary and pulsational parameters, we combined \ph\ with {\sc emcee}, a {\sc python} implementation of the affine invariant ensemble sampler for Markov chain Monte Carlo (\mcmc) proposed by \citet{Goodman2010} and implemented by \citet{DFM2013}.

\begin{figure*} 
\includegraphics[width=14cm]{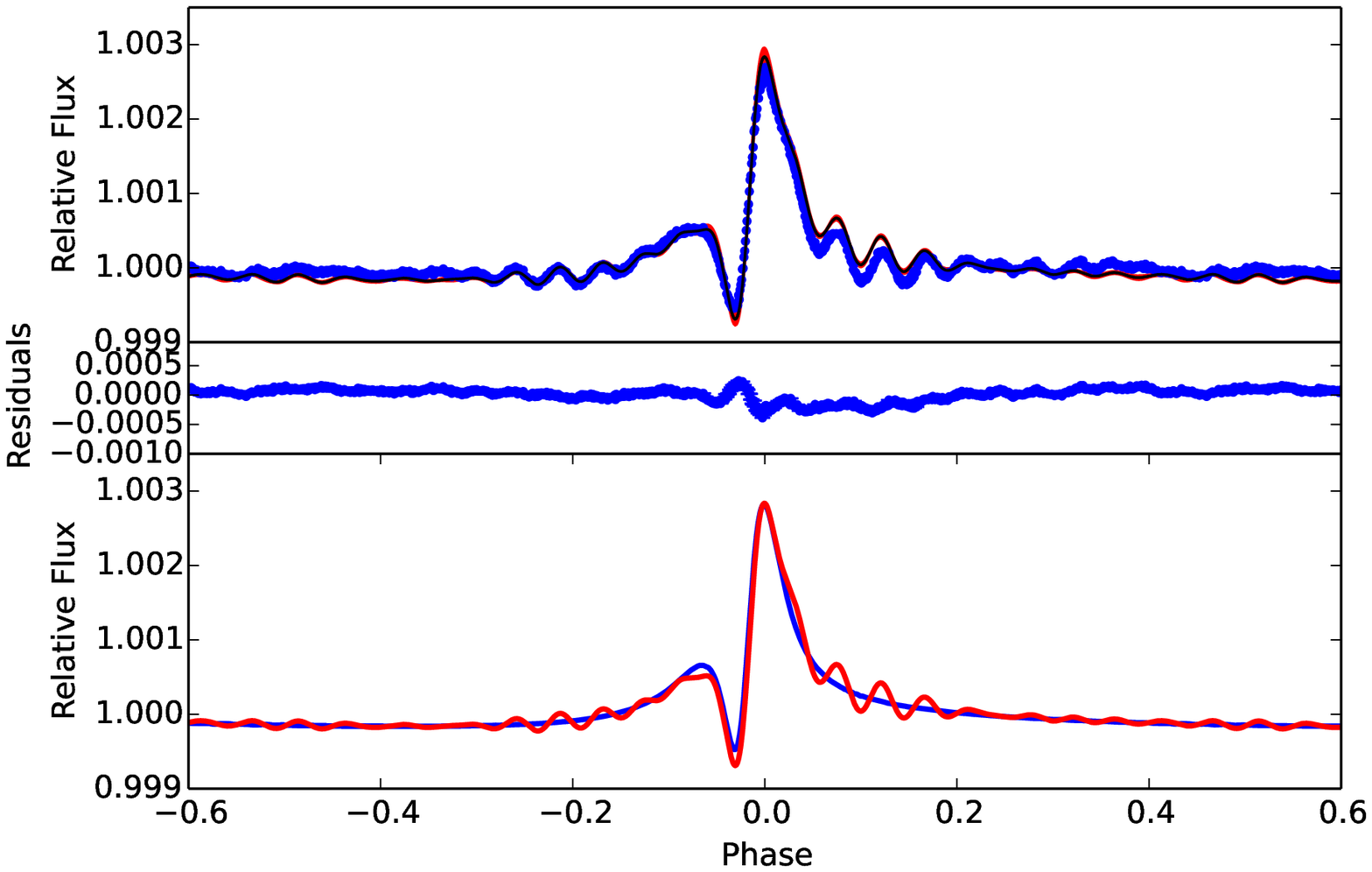}  
\vspace{-1cm}
\small\caption{Upper panel: Best-fit \ph\ model including 7 pulsations (black line). The observed light curve (blue points) was prepared as specified in \S\ref{sec:binary}. The red envelope depicts the 1\,$\sigma$ spread of the all the models in the final mcmc iteration (128 models), which were determined using \mcmc. Middle panel: the residuals (blue points) of the best-fit model. Lower panel: a comparison of the theoretical model with 7 pulsations (red) and the theoretical model with no pulsations (blue).}
\label{fig:lc_model} 
\end{figure*} 

\begin{figure*}
\hfill{}
\includegraphics[width=\hsize, height=10cm]{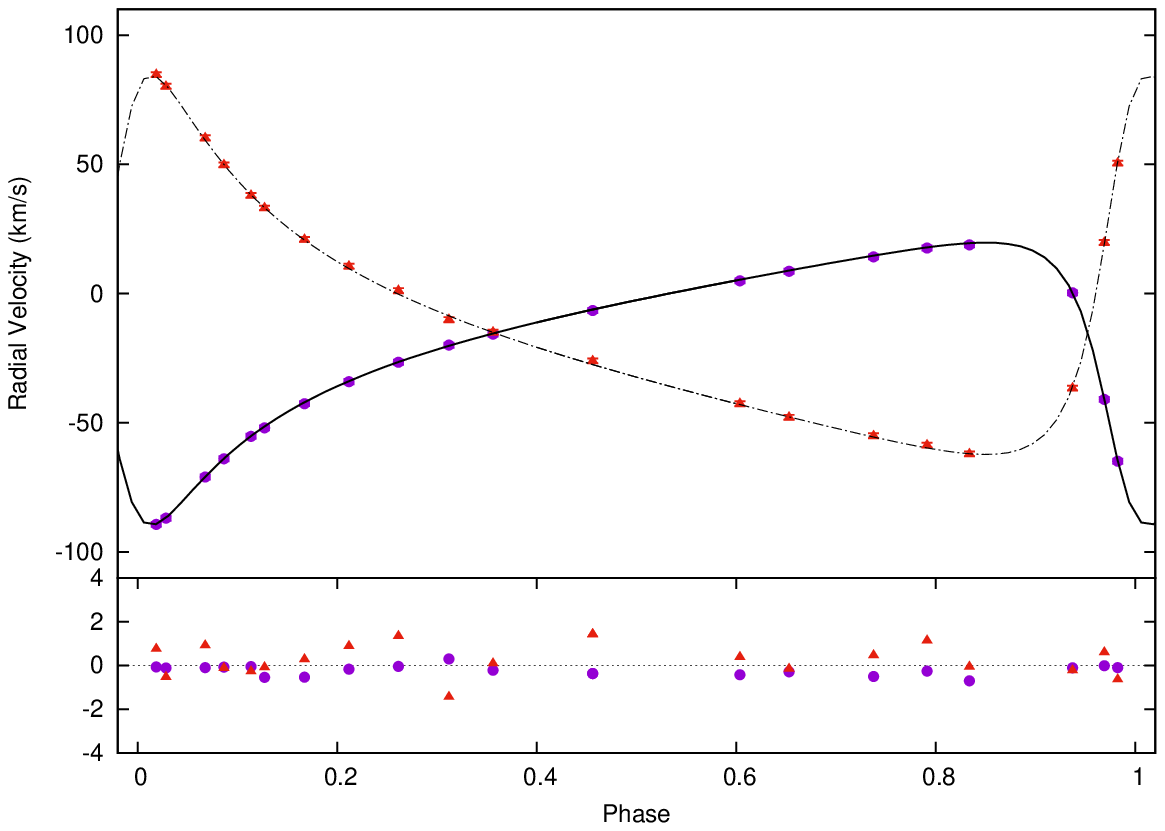}\\
\small\caption{Upper panel: the observed radial velocities of KIC\,3749404 for the primary (purple circles) and secondary component (red triangles). The uncertainties are plotted but barely visible in comparison with the point sizes. The best-fit model (where the light curve model includes 7 pulsations) is shown for the primary component (solid line) and secondary component (dashed line). Lower panel: the residuals to the best fit model. A dotted horizontal line is placed at zero for clarity. See \S\ref{sec:mod} for details of the binary star model. The spectra from which these radial velocities were derived were observed between UT 2011 May 10 and 2011 Jun 23, using the Tillinghast Reflector Echelle Spectrograph (TRES) mounted on the 1.5-m Tillinghast Reflector at the Fred L. Whipple Observatory on Mt. Hopkins, AZ.}
\label{fig:rvs}
\hfill{}
\end{figure*}

\mcmc\ explores the binary parameter space using a set of Markov chains, in this case 128. These chains begin with random distributions of each parameter based only on the prior probability distributions. They move through parameter space by assessing their posteriors at each point and then selecting a new position based on the position of another chain. The step size is based on the covariance of the two chains. If the move increases the a posterior likelihood then it is accepted; if the move decreases it then it may be accepted with a certain probability. During the initial burn-in time the Markov chains are merging towards their maximum likelihood position. The statistics of a large number of iterations provide posterior distributions for the model parameters. 

We generated two models for KIC\,3749404, one containing the most significant pulsations, and one without pulsations (to see the effect of pulsations on the binary star parameters). We used over 20\,000 iterations to arrive at each model and determined that this was sufficient by assessing the change in log likelihood and the stability of each parameter as a function of iteration (i.e. all the chains were required to plateau for all parameters). The model without pulsations comprises 10 binary star parameters. From the residuals of this model, we identified seven pulsations with amplitudes greater than 20\mumag, thus the second model contains 7 pulsations and comprises 24 parameters: 10 binary star parameters and an amplitude and phase parameter for each pulsation. As the fitted pulsations are tidally induced, we fixed each pulsation frequency to a multiple of the orbital frequency (as determined from our frequency analysis). The pulsations were modelled by simultaneously fitting sine waves with the binary star model. 

The binary parameters were selected based on their contribution to the observed flux variation to KIC\,3749404. For the results of the best-fit models, see Table\,\ref{tab:ParamFree} for a list of fundamental parameters and Table\,\ref{tab:freqs} for a list of the pulsation parameters from the model with 7 pulsations. For the binary star model, the inclination, eccentricity, argument of periastron, phase shift, primary and secondary radii, and primary and secondary gravity darkening exponents were sampled using \mcmc. The phase shift is the horizontal offset required to keep the model's periastron variation centred at phase 0.0 when changing the argument of periastron and eccentricity in the phase folded data. 

For each model iteration the limb darkening exponents were calculated; the luminosity was fitted using least-squares; and the effect of Doppler boosting was added to the light curve for each component. Doppler boosting is caused by the radial motion of the two stars and is the combined effect of the Doppler shifting of the stars' spectral energy distributions,
aberration and an altered photon arrival rate. The net result of Doppler boosting is an increase in the observed flux from a star when it moves towards the observer, and a decrease when it moves away. It was predicted by \citet{Loeb2003} and \citet{Zucker2007b}, and has recently been observed in several systems from ground-based data as well as \textit{Kepler}\ and CoRoT light curves \citep[see e.g.][]{Mazeh2010,van-Kerkwijk2010,Shporer2010,Bloemen2011}. To model the Doppler boosting signal, Eqn\,2 in \citet{Bloemen2011} provides a function that can easily be applied to the binary star model:

\begin{equation}
F_{\lambda} = F_{0,\lambda} \left[R\left(1 - B_1\frac{v_{r,1}}{c}\right) - \left(1-R\right)\left(B_2\frac{v_{r,2}}{c}\right)\right],
\end{equation}

\noindent where $F_{\lambda}$ is the observed flux, $F_{0,\lambda}$ is the emitted flux, $R$ is the light ratio, $v_{r,1}$ and $v_{r,2}$ are the radial velocities of the primary and secondary components, $B_1$ and $B_2$ are the passband-weighted boosting factors, where $B=5+\rm{d}\ln F_{\lambda}/\rm{d}ln\lambda$ \citep{Loeb2003}.

For each parameter we used a flat, uniform prior. The prior ranges were selected to encompass all physical models given the spectroscopic information. We restricted the prior on the inclination to be below 90\degs to avoid obtaining a double peaked distribution reflected about 90\degs. We also restricted the gravity darkening exponents to the ranges $0.8 -1.0$ for the primary and $0.2 - 0.6$ for the secondary component, where the gravity darkening exponents are predicted to be 1.0 and 0.32 for stars with radiative \citep{vonZeipel1924} and convective \citep{Lucy1968} envelopes, respectively. 

In our model we assumed that the noise is Gaussian, which does not take into account correlated noise contributions (see, e.g., Barclay et al. 2015)\nocite{Barclay2015}. Consequently, it is likely our uncertainties are underestimated. Fig.\,\ref{fig:posteriors} shows the posterior distributions of the best-fit model. It can be seen that all the parameter histograms (top left to bottom right diagonal plots) are normal distributions, with the exception of the secondary gravity brightening exponent, which shows a slightly skewed distribution. The upper panel of Fig.\,\ref{fig:lc_model} shows the best-fit \ph\ model with 7 pulsations (black with red, 1\,$\sigma$ envelope) and the best-fit \ph\ model with no pulsations included (green). The lower panel shows the residuals from the model including pulsations. Both models are the average of all the lightcurve models from the last iteration (128), one from each Markov chain. The envelope denotes the 1-$\sigma$ spread of results determined by finding the standard deviation of the light curve models. Fig.\,\ref{fig:rvs} depicts the best-fit model to the radial velocity data for the \ph\ model with 7 pulsations, which was fitted simultaneously with the light curve. 

In the light curve, it can be seen that there is a significant deviation of the model from the data in the phases following the periastron variation and in the regions that are flatter, relative to the periastron variation. The cause of this may lie in detrending, as \hb\ star light curves often do not have an obvious baseline. All remaining features in the observed data are fit adequately by the model. The more pronounced red regions near the peak and trough of the periastron variation show that the models are sampling a wider distribution in these areas, which are encapsulated in the uncertainties.

When comparing the two models, the most striking differences stem from different inclinations of the two models: $i = 60.2 \pm 1.0$\degs and $i = 62 \pm 1.0$\degs\ for the models with zero and seven pulsations, respectively. These differences (although only at the 2$\sigma$ level) translate to a significantly larger semi-major axis in the model with pulsations (2.7\,\Rsun\ larger), and larger masses  (0.28\,$\Msun$ for the primary and 0.12\,$\Msun$ for the secondary; see Table~4), with the pulsation model favouring the more massive stars. Finally, the radii of both components are larger for the model with pulsations. These differences convey the importance of modelling pulsations simultaneously with the light curve.

\begin{table}
\caption{The pulsation amplitude and phase values for the binary model combined with tidally induced pulsations. The frequencies were fixed to precise multiples (harmonics) of the orbital frequency. The phase is relative to the binary model zero point in time.
\label{tab:freqs}
}
\begin{center} 
\begin{tabular}{ccc r} 
\hline 
Multiple of & Frequency & Amplitude & \multicolumn{1}{c}{phase}\\
$\nu_{orb}$ &  \perday & ppm & \multicolumn{1}{c}{rad}\\
\hline
18& 0.8864(6)&18(2)&5.0(1)\\
20& 0.9849(7)&39(4)&3.05(9)\\
21& 1.0341(7)&52(8)&2.50(7)\\
22& 1.0834(7)&22(5)&2.4(1)\\
23& 1.1326(7)&36(7)&1.06(8)\\
24& 1.1818(7)&60(5)&0.90(8)\\
25& 1.2311(8)&15(6)&0.90(8)\\
\hline
\end{tabular}
\end{center} 
\end{table} 

\subsection{Spectroscopic Analysis and the Case for the Am Nature of the Primary Component \label{sec:am}}

To determine the spectroscopic stellar parameters, we performed an
analysis similar to those used to characterize the stars of the
circumbinary planet-hosting binary systems Kepler-34, Kepler-35, and
KOI-2939 \citep{Welsh2012,Kostov2015}, though the properties of
KIC\,3749404 necessitated a few changes. Blended lines can prevent accurate classifications, so we included in this analysis only the $14$\ TRES spectra that have a velocity separation greater
than $40$\,$\kms$\ between the two stars. We began by cross-correlating
the TRES spectra against a five-dimensional grid of synthetic
composite spectra. The grid we used for KIC\,3749404 contains every
combination of stellar parameters in the ranges $T_1=[5500,8500]$,
$T_2=[5500,8500]$, $\log{g_1}=[2.5,5.0]$, $\log{g_2}=[3.0,5.0]$, and
${\rm [M/H]}=[-1.0,+0.5]$, with grid spacings of $250$\,K in $T$, and
$0.5$\,dex in $\log{g}$\ and in [M/H] (20,280 total grid
points).\footnote{We ran a separate TODCOR grid solely to determine
the $v\sin{i}$\ values, which we left fixed in the larger grid. This
is justified because the magnitude of the covariance between
$v\sin{i}$\ and the other parameters is small. This simplification
reduces computation time by almost two orders of magnitude. The final $v\sin{i}$ values are $v_1\sin i = 29(2) \kms$ and $v_2\sin i = 9(2) \kms$.} At each
step in the grid, TODCOR was run in order to determine the radial velocities of the
two stars and the light ratio that produces the best-fit set of $14$\
synthetic composite spectra, and we saved the resulting mean
correlation peak height from these $14$\ correlations. Finally, we
interpolated along the grid surface defined by these peak heights to
arrive at the best-fit combination of stellar parameters. 

The spectroscopic analysis was limited by the degeneracy among the parameters (i.e., a nearly equally good fit was obtained by
slightly increasing or decreasing $T$, $\log{g}$, and [M/H] in
tandem), but the light curve model provided independently determined
surface gravities (see \S\ref{sec:binary}) that were partially able to
lift the degeneracy. We thus included these gravities in our
analysis, and found that the best fit occurred for a very high
metallicity (${\rm [M/H]}>+0.5$); however the resulting parameters and light ratio of the spectral templates were inconsistent with the binary model. The expected flux ratio can be approximated by assuming black body radiation and integrating Planck's law across the TODCOR
bandpass:
\begin{equation}
\frac{F_2}{F_1} = \left(\frac{R_2}{R_1}\right)^2 \int_{\lambda_i}^{\lambda_f}\frac{e^{hc/(\lambda k_B T_1)} - 1}{e^{hc/(\lambda k_B T_2)} - 1}\,d\lambda.
\end{equation}
Recognizing that the combination of surface gravities with the mass
ratio from the radial velocities can yield the area ratio, the above equation can be
rewritten solely in terms of the spectroscopic observables:
\begin{equation}
\frac{F_2}{F_1} = q \left(10^{\log{g_1} - \log{g_2}}\right) \int_{\lambda_i}^{\lambda_f}\frac{e^{hc/(\lambda k_B T_1)} - 1}{e^{hc/(\lambda k_B T_2)} - 1}\,d\lambda,
\end{equation}
where $q$\ is the mass ratio, $M_2/M_1$. For our best-fit templates,
the measured and expected light ratios differed by more than a factor of
$2$, at high significance. 

These inconsistencies in the light ratio and the
temperatures indicated that the spectroscopic model was not yet
sufficient. Given that the primary is an A star and the best fit occurred for high metallicities, we next explored the possibility that the two stars have different apparent metallicities. If the primary is
an Am star, displaying photospheric enrichment in metallic lines, we
would expect the spectra for the primary and secondary to be fit by
templates of different metallicity even though they presumably formed
with identical compositions.

We first tested this by running new grids of TODCOR correlations that
include a sixth dimension, along which lies the metallicity of the
secondary. This improved the agreement between the spectral templates and binary star model, including the light ratios, which were then within 25\% of each other. The newly determined metallicities of the primary and secondary components were ${\rm [m_1/H]}>+0.50$ and ${\rm [m_2/H]}=0.00$, respectively. While we are only able to report lower limits for the primary because our library of synthetic spectra only includes metallicities up to ${\rm [m/H]}=+0.5$, these
results suggest that the primary may be an Am star (and that the secondary is not). 

Consequently, we examined spectra taken at quadrature for the chemical abundance anomalies seen in Am stars. The observed spectrum was compared with the synthesized binary spectra using the algorithm developed and described by \citet{Murphy2015}. Suitable atmospheric parameters for the components of the binary system were chosen such that the synthetic hydrogen line profiles matched the observed profiles. A satisfactory match was obtained with $T_1 = 8000(300)$\,K, $T_2 = 6900(300)$\,K, $\log g_1 = 3.8(3)$, $\log g_2 = 4.0(3)$, $v_1\sin i = 30(5)$\,km\,s$^{-1}$ and $v_2\sin i = 10(5)$\,km\,s$^{-1}$, using $[$Fe/H$]=0.0$ for both components. This configuration leads to a mass ratio of $M_2/M_1 = 0.74$ and a light ratio of $L_1/(L_1 + L_2) = 0.80$. The synthetic and observed spectra were smoothed to classification resolution (2.5\,\AA\ per 2 pixel) to reduce the effects of noise and additional sources of spectral line broadening that are hard to quantify \citep{Murphy2016}. It is worth noting that, due to the applied smoothing, the $v \sin i$ values of the initial analysis are more reliable ($v_1\sin i = 29 (2) \kms$ and $v_2\sin i = 9(2) \kms$).  With these parameters, the observed metal line profiles were much stronger than the synthetic ones. By increasing the {\it global} metallicity of the primary component, the fit to the metal lines improved at the expense of the fit to the hydrogen lines; increasing by $+0.5$\,dex, the fit to the hydrogen line profiles worsened considerably, with substantial further improvement to the metal lines required. 

For the primary component, agreement with the hydrogen line profiles could have been restored with small changes to $T_1$ and $\log g_1$, but the morphology of the metal lines could not be improved with further changes to the global metallicity, alone. Already at $+0.2$\,dex, the synthetic Ca II K line was stronger than the observed spectrum, while the Fe, Sr and Ti lines remained poorly matched. Increasing the metallicity to $+0.5$\,dex resulted in discrepant Ca II K and hydrogen lines, and substantial further enhancement of the synthetic Sr and Fe lines was still required. Thus the observations show selective metal enrichment of the photosphere, which match the abundance patterns of Am stars (see, e.g. \citealt{Murphy2012}). 

A-type stars in close binary systems commonly show Am peculiarities (\citealt{Abt1967}; see \citealt{Murphy2014b} for a review). The fact that the primary is an Am star is therefore expected. The secondary star is an early F star for which Am-type peculiarity is less common \citep{Wolff1983}, presumably due to the increase in convective mixing. The derivation of abundances for a chemically peculiar star in a double-lined spectroscopic binary system is a state-of-the-art challenge beyond the scope of this paper. For the purpose of this work, our current analysis suggests that the appearance of the spectra is consistent with the light curve model, which we described in detail in \S\ref{sec:binary}.

\section{Apsidal Motion \label{sec:apsmo}}

\begin{figure*}
\centering
\hspace{-0.5cm}
\includegraphics[height=1.0\textwidth,angle=270]{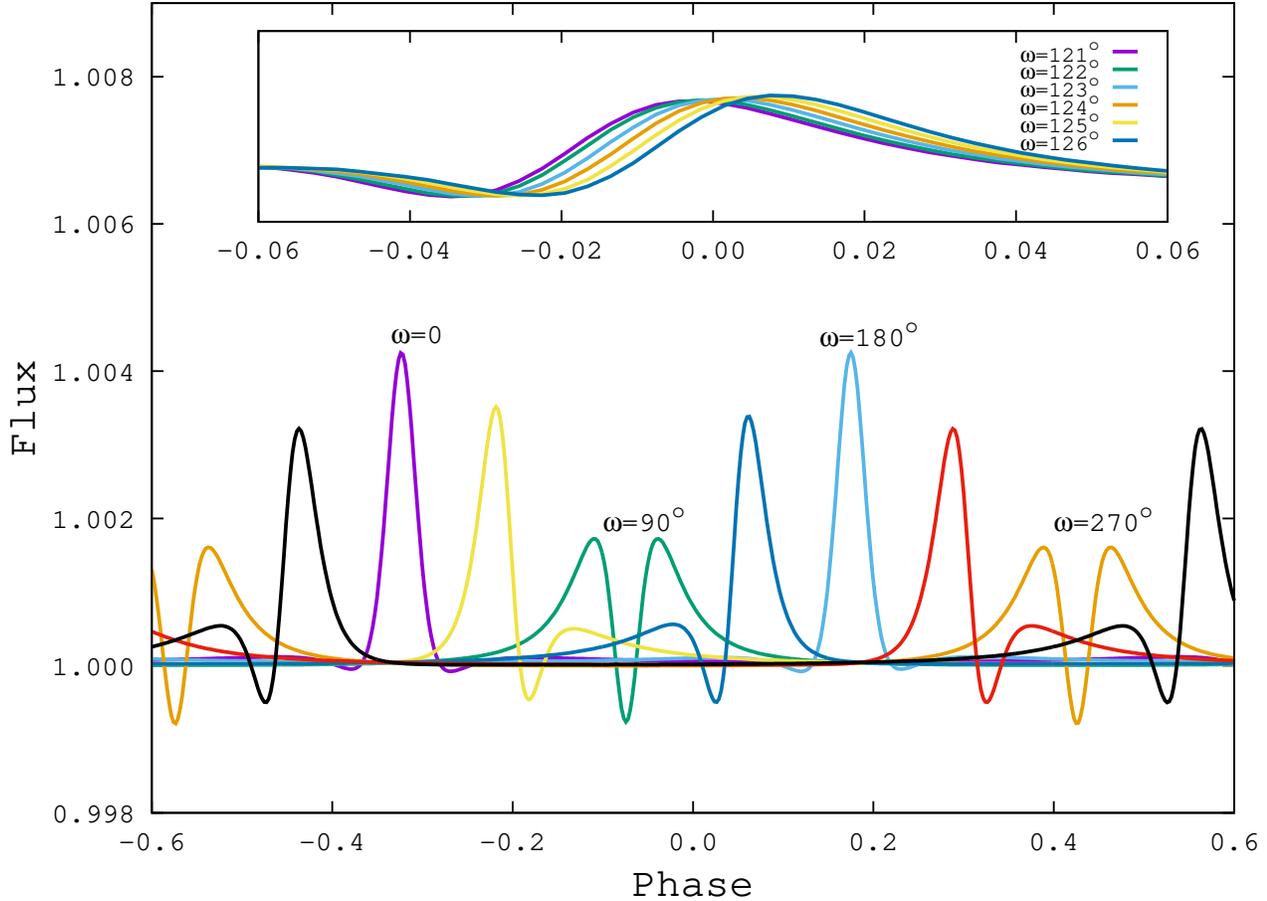}  
\bigskip
\small\caption{Models of KIC\,3749404 with different values of $\omega$, the argument of periastron. The figure highlights the effect of the periastron value on the shape of the periastron variation for one full orbital cycle about the centre of mass. The periastron variation drastically changes both its position in phase space and its shape (width and amplitude) with the changing values of $\omega$. The inset shows the measured effect on the light curve shape over the 4-yr length of the Kepler data set used. \label{fig:apsmo} }
\end{figure*} 

Apsidal motion is the rotation of the elliptical orbit about the centre of mass. The rate of classical apsidal advance is determined by the non-spherical mass distribution within the stars due to tidal distortions and rotation. General relativity produces additional apsidal motion. The net apsidal motion depends on the component's masses, radii, rotation rates, and density profiles, in addition the orbital semi-major axis and eccentricity.

When initially phasing the 4-yr light curve of KIC\,3749404, it was clear, due to the significant smearing of the periastron variation, that KIC\,3749404 is undergoing rapid apsidal motion. Fig.\,\ref{fig:apsmo} depicts several models showing the change of the periastron variation of KIC\,3749404 as a function of the argument of periastron for a complete cycle about the centre of mass. The insert shows a model depicting the apsidal advance over the 4\,yr data set. 

To determine the rate of periastron advance, we fitted $\omega$ at the beginning (Quarters 1 and 2, LC data) and end (Quarters 15 and 16, LC data) of our data set (Quarters 0 and 17 are short Quarters that did not contain any periastron variations and thus were not used). As our aim was to determine the change in the argument of periastron, we fixed all parameters except the argument of periastron and phase shift, to those determined for the best-fit model of Quarters 8--10 (including pulsations). We also included a global pulsation phase-shift (which shifts all the pulsations by the same amount), as the phases of the pulsations were observed to shift by a different amount than the longitude of periastron. Since the pulsation phases depends on the azimuthal numbers of the tidally excited modes, measuring this phase shift can help identify them.

\begin{figure*}
\centering
\begin{minipage}[b]{0.5\linewidth}
\hspace{1.8cm}
\includegraphics[width=7cm]{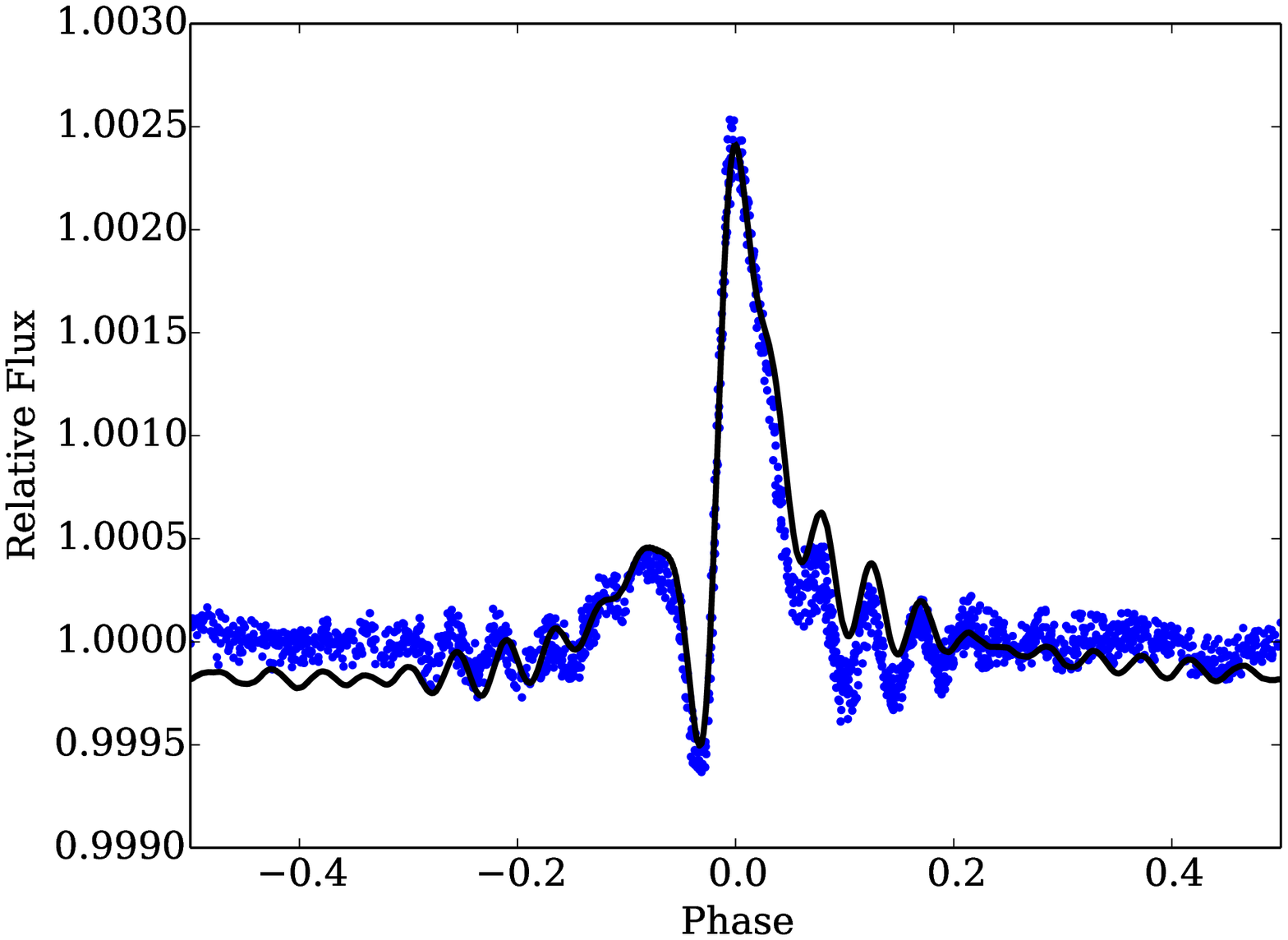}
\end{minipage}%
\begin{minipage}[b]{0.5\linewidth}
\includegraphics[width=7cm]{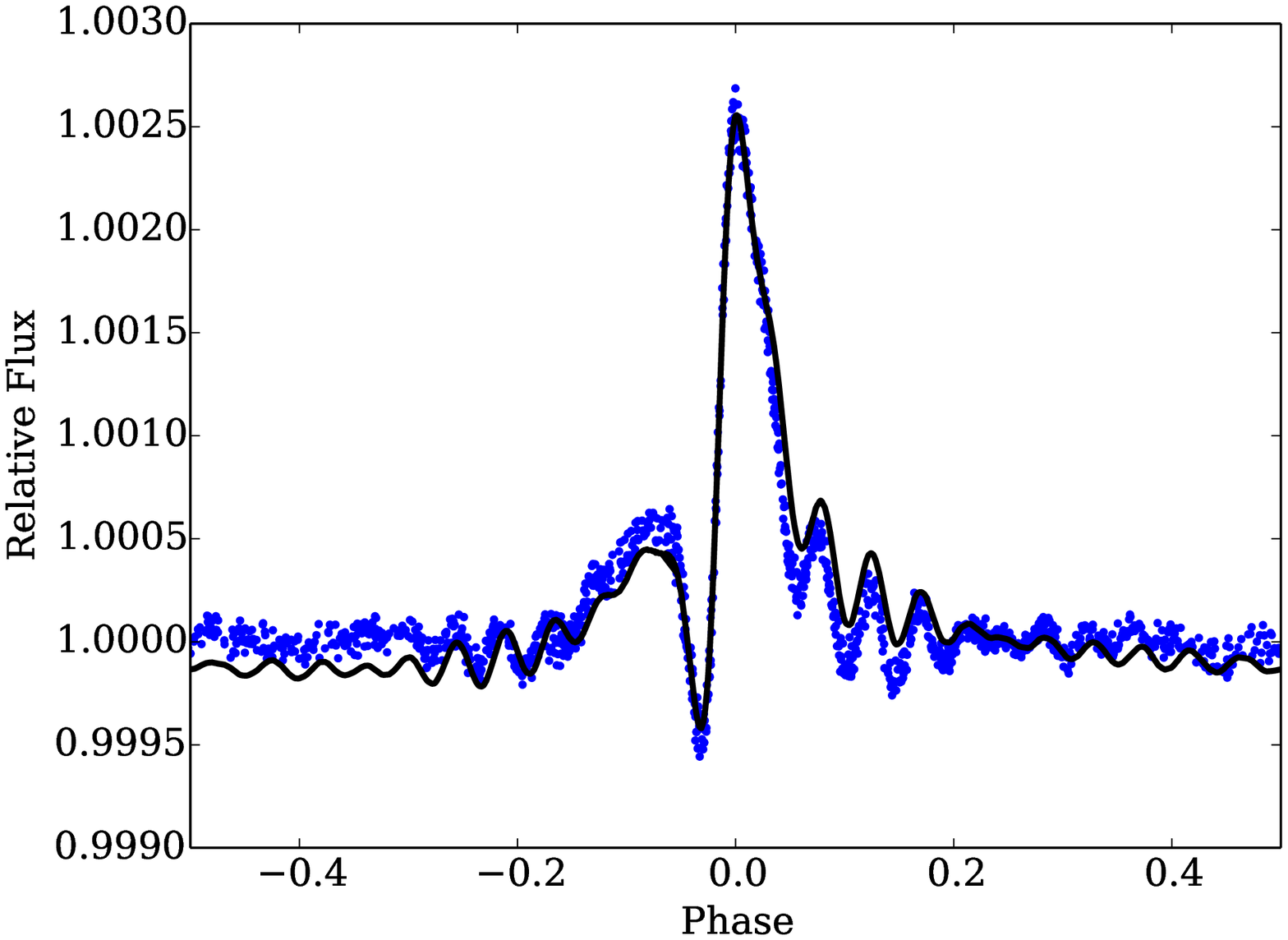}
\end{minipage}\\%
  \centering
  \includegraphics[width=0.8\linewidth]{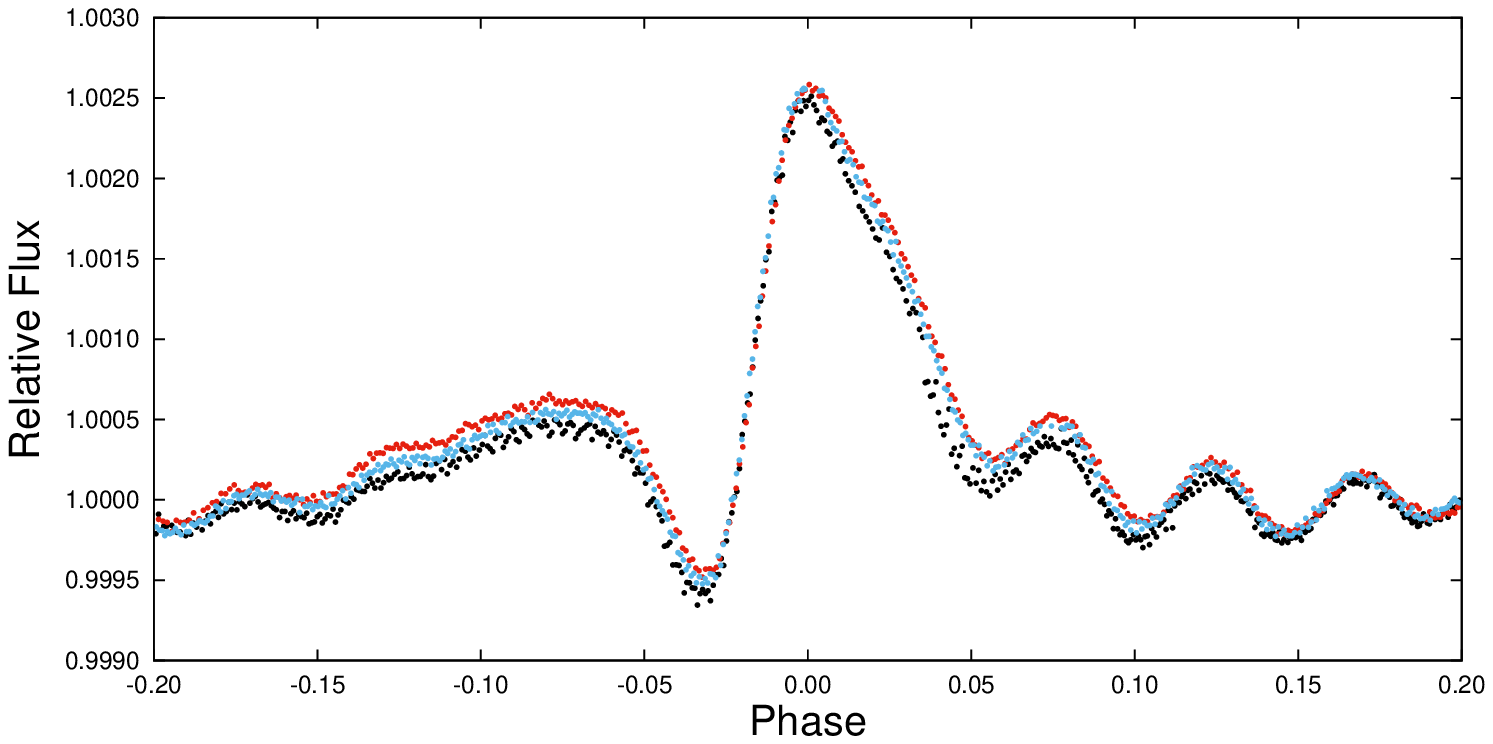}
\caption{Top panel: the best-fit models (black lines) for the initial segment of data (left panel) of Quarters 1 and 2 (blue dots), and the final segment of data (right panel) for Quarters 15 and 16 (blue dots). To perform the fit, the parameters from the model with pulsations were fixed, with the exception of the argument of periastron, phase shift and a global pulsation phase shift parameter, which were fitted. Bottom panel: The binned, phase-folded \kep\ data of Quarters 1 and 2 (black); Quarters 8, 9 and 10 (blue); and Quarters 15 and 16 (red) between phases -0.2 and 0.2. The data clearly demonstrate the changing shape of the periastron variation due to the advance of periastron.}
\label{fig:apsmo_model}
\end{figure*}

\begin{figure}
\centering
\includegraphics[width=8cm]{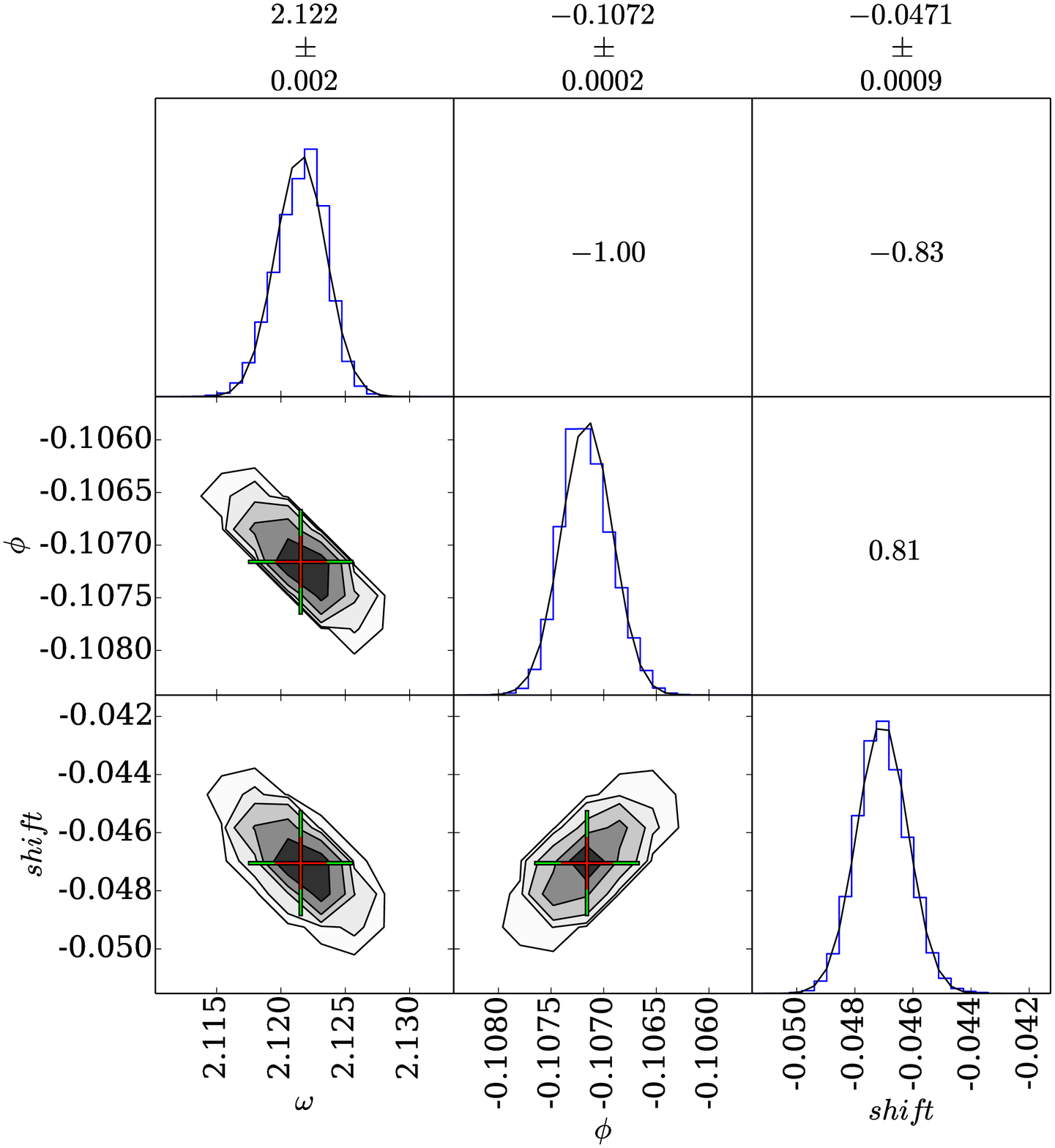}
\includegraphics[width=8cm]{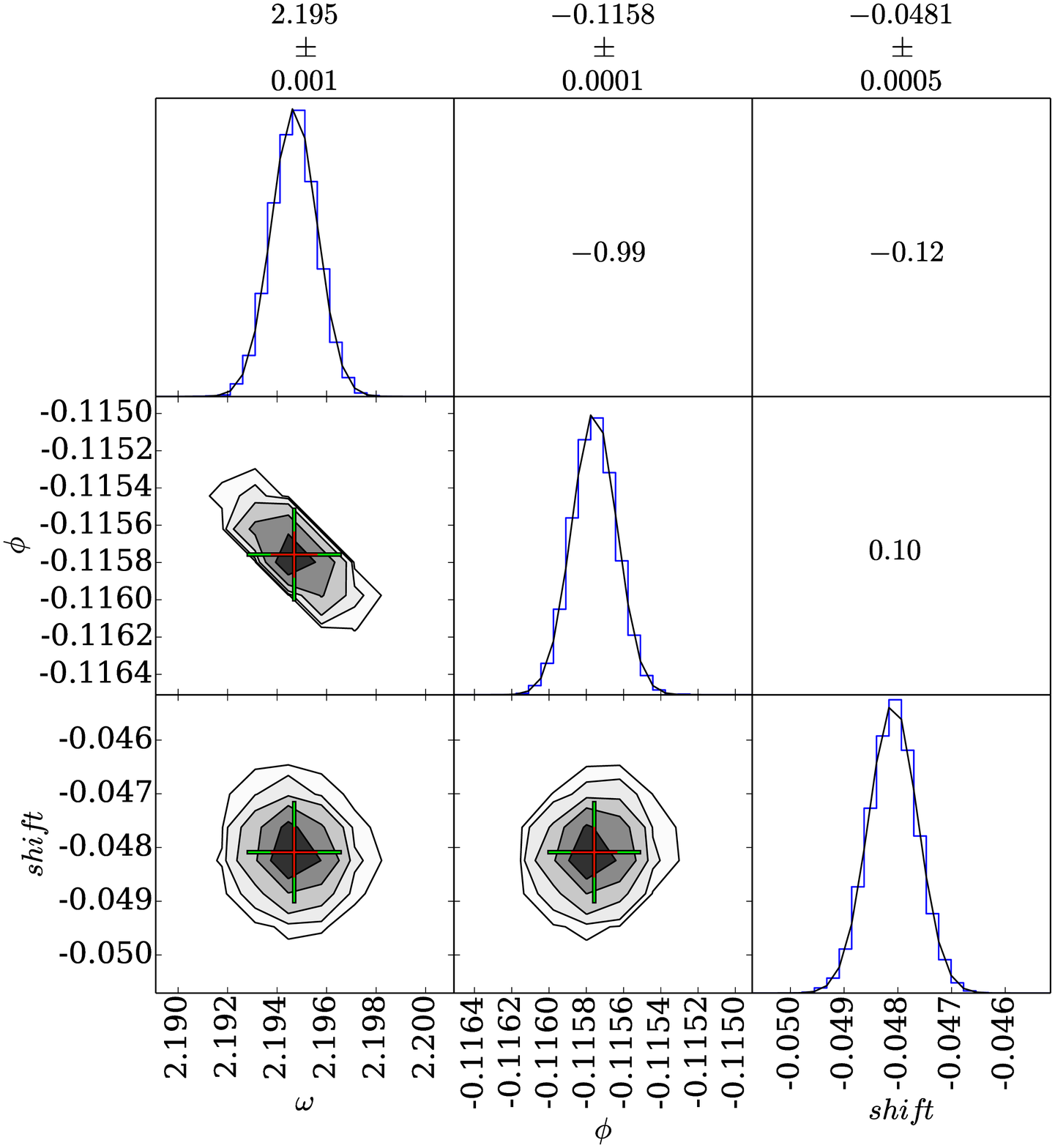}
\caption{Posterior distributions of the apsidal motion parameters for the model of the first section (top panel) and last section (bottom panel) of \kep\ data. Here $\omega$ is the argument of periastron; $\phi$ is the light curve phase shift, which is required to keep the periastron variation centred when changing the argument of periastron; and $shift$ represents the amount by which all the pulsation phases are shifted with respect to their values for the original model with 7 pulsations. The layout is analogous to that of Fig.\,\ref{fig:posteriors}.}
\label{fig:apsmo_posteriors}
\end{figure}

We selected the initial and final data sets so that they contained five orbits. Using this method we determined the initial and final values for the argument of periastron to be 2.122(2) and 2.195(1), respectively (see Figs\,\ref{fig:apsmo_model} and \ref{fig:apsmo_posteriors} for the model fits and posteriors, respectively). It is important to note that we did not account for silght changes in the inclination or individual pulsation phases and as such our uncertainties are likely underestimated. For the two data sets we selected the peak of the periastron variation in the middle of the data set as the zero point in time for that section of data. To obtain the rate of periastron advance we divided the change in the argument of periastron by the difference in time between the two data sets (1309.773(4)\,d). We determined the rate of apsidal advance to be 1.166(1)\degs/year. To ensure that changes in the light curve were not caused by the changing amplitudes of the tidally induced pulsations, we applied a Fourier transform to the beginning and end segments of data to compare the amplitudes. The mode amplitudes were found to be constant at the ~1 ppm level, while the light curve changes at the ~100 ppm level over the duration of the Kepler data set (see Fig.\,\ref{fig:apsmo_model}).

To compare our observed value for the rate of apsidal advance with the predicted rate, we then used the tables of \citet{Claret1997} to obtain the apsidal motion constants (Love numbers) of the primary and secondary components to be $k_{2,1} = 0.0036(5)$ and $k_{2,2} = 0.0042(4)$, for models similar to the stars in KIC\,3749404 (where $k_2$\,=\,0 for a point mass and $k$$_2$\,=\,0.75 for a homogeneous sphere). We assumed that both components of KIC\,3749404 are main-sequence stars and used the fundamental parameters from the model with pulsations. Using these values we calculated the classical theoretical rate of apsidal advance \citep{Cowling1938,Sterne1939,Kopal1959} using: 

\begin{equation}
\begin{split}
\dot{\omega}_{CL}^{theor}(deg/yr) = \ & 365.25\left(\frac{360}{P}\right)\bigg\{k_{2,1}r^5_1\bigg[15f_2(e)(M_2/M_1) 
\\&+ \left(\frac{\widetilde{\omega}_{r,1}}{\widetilde{\omega}_k}\right)^2\left(\frac{1+M_2/M_1}{(1-e^2)^2}\right)\bigg] 
\\&+ k_{2,2}r^5_2\bigg[15f_2(e)(M_1/M_2) 
\\&+ \left(\frac{\widetilde{\omega}_{r,2}}{\widetilde{\omega}_k}\right)^2\left(\frac{1+M_1/M_2}{(1-e^2)^2}\right)\bigg]\bigg\},
\end{split}
\end{equation}

\noindent where $P$ is the orbital period, $k_{2,1}$ and $k_{2,2}$ are the apsidal motion constants for the primary and secondary components, respectively, $f_2(e) = (1 + 3/2e^2 + 1/8 e^4)(1-e^2)^{-5}$ where $e$ is the eccentricity, $M_1$ and $M_2$ are the masses of the primary and secondary components in solar mass units, $r_1$ and $r_2$ are the radii of the primary and secondary component in terms of the semi-major axis, $\widetilde{\omega}_{r,1}$ and $\widetilde{\omega}_{r,2}$ are the angular axial rotational speeds of the primary and secondary stars, and $\widetilde{\omega}_k = 2\pi/P$. For the classical rate of apsidal advance we obtained $\dot{\omega}_{CL}^{theor} = 0.002(7)$\degs/yr. As KIC\,3749404 has a large eccentricity and relatively short orbital period, the general relativistic contribution is significant. Thus we calculated the general relativistic apsidal motion term \citep{Levi-civita1937,Kopal1959}:

\begin{equation}
\dot{\omega}_{GR}^{theor}(deg/yr) = 9.2872 \times 10^{-3}\frac{(M_1+M_2)^{2/3}}{(P/2\pi)^{5/3}(1-e^2)}(deg/yr).
\end{equation}

\noindent We obtained the general relativistic contribution to be $\dot{\omega}_{GR}^{theor} = 0.005(1)$\degs/yr giving a combined value of $\dot{\omega}_{CL+GR}^{theor} = 0.007(6)$\degs/yr.

The observed rate of apsidal advance of 1.166(1)\degs/yr determined through light curve modelling is two orders of magnitude larger than the theoretically predicted rate of 0.007(6)\degs/yr. We are aware that the classical, theoretical rate of apsidal advance is strongly dependent on the stellar radii (to the 5$^{th}$ power) and that, as the light curve of KIC\,3749404 does not contain eclipses, our radii determination is based on the Roche lobe geometry (which is not as accurate as direct detection through eclipse modelling). However, the difference between the theoretical and predicted rates is highly significant (even when accounting for slightly underestimated uncertainties) and, consequently, there is no way that the apsidal motion rate of KIC\,3749404 could be reduced to the theoretical rate through tweaking assumptions or inflating uncertainties. We hypothesise that the rapid rate of apsidal advance of KIC\,3749404 is due to a tertiary component in the system.

\subsection{Precession from Tidally Induced Pulsations}

In addition to apsidal motion produced by the rotational and equilibrium tidal distortion of the stars, the dynamical tidal distortion (i.e., tidally induced pulsations) can produce significant orbital precession. Tidally induced pulsations are typically produced by gravity modes within the stellar components, which create oscillations in the gravitational fields of the stars, in addition to the luminosity variations seen in Fig.\,\ref{fig:lc_model}. The aspherically distorted gravitational field (averaged over an orbital cycle) can produce significant orbital precession, depending on the amplitude to which the gravity modes are excited. 

The effect of dynamical tides on apsidal motion has previously been investigated in \citet{Smeyers2001, Willems2002, Claret2002} for binaries of various stellar masses, orbital periods, and eccentricities. These authors found that the contribution of dynamical tides to orbital precession is generally small (especially at orbital periods exceeding 5 days), except when very near resonance with stellar oscillation modes. Even in this case, the precession rate is typically only altered at a level of order unity, and it cannot be enhanced by a factor of $\sim 100$ as is required to explain the apsidal motion in KIC\,3749404. Moreover, non of the tidally excited gravity modes in the components of KIC\,3749404 are particularly close to resonance. We infer this from comparison with other \hb\ stars that have pulsations near resonance (e.g., KOI-54, \citet{Welsh2011}; KIC\,8164262, Hambleton et al., in prep) where the pulsations close to resonance have significantly larger amplitudes than other pulsations in the system. Therefore, we find it unlikely that tidally induced pulsations can produce the rapid apsidal motion of KIC\,3749404.

\subsection{Precession from a Third Body}

One possibility is that the rapid apsidal motion in KIC\,3749404 is caused by the gravitational influence of an external perturber. To order of magnitude, the precession rate due to an external third body is \citep{Eggleton2001}:
\begin{equation}
\label{omega1}
\dot{\omega}_{3B} \approx \frac{M_3}{M_1+M_2+M_3} \frac{1}{(1-e_{\rm in}^2)^{1/2} (1-e_{\rm out}^2)^{3/2}} \frac{\Omega_{\rm out}}{\Omega_{\rm in}} \Omega_{\rm out} \, .
\end{equation}
Here, $M_3$ is the mass of the third body, while $e_{\rm in}$ and $e_{\rm out}$ refer to the eccentricities of the inner and outer orbits, and $\Omega_{\rm in}$ and $\Omega_{\rm out}$ refer to their orbital angular frequencies. The precise precession rate depends on the relative inclination of the systems which is not constrained.

We expect the third body to have $M_3 < M_1,M_2$ because it is not visible in the spectra, and orbital stability requires $\Omega_{\rm out} \ll \Omega_{\rm in}$. Making the simplifying assumption of $e_{\rm out}=0$, and using the measured values of $e_{\rm in}$ and $\Omega_{\rm in}$ yields
\begin{equation}
\label{omega2}
\dot{\omega}_{3B} \sim \frac{M_3}{M_1+M_2+M_3} \bigg(\frac{P_{\rm in}}{P_{\rm out}}\bigg)^2 10^4 \, {\rm deg}/{\rm yr} \, .
\end{equation}
Equating this with the observed rate of precession of $\dot{\omega}_{\rm obs} \simeq 1 \, {\rm deg}/{\rm yr}$ yields the requirement
\begin{equation}
\label{omega3}
\frac{M_3}{M_1+M_2+M_3} \bigg(\frac{P_{\rm in}}{P_{\rm out}}\bigg)^2 \sim 10^{-4} \, .
\end{equation}

A typical low-mass companion of $M_3 \sim 0.5 \, M_\odot$ would then require $P_{\rm out} \sim 2 \, {\rm yr}$. Of course, different companion masses, eccentricities, and inclinations would change the required orbital period of the third body, but we expect this period to be on the order of years to produce the observed apsidal motion. A third body at this orbital separation, inclination and eccentricity would have a semi-amplitude of $\sim$5\,km\,s$^{-1}$ and could likely be detected through long term RV variations of the primary binary, or possibly via an infrared excess in the spectrum. We encourage follow-up observations of KIC\,3749404 to constrain the nature of a putative third body and determine whether the observed precession is driven by three-body processes or tidal interactions.

Dependent on the inclination of the tertiary component's orbit, it is also possible that nodal precession could occur. Given a long enough time base, this could also be detected in the orbit as a change in the radial velocity amplitude due to the motion of the inner binaries center of mass (see \eg\ \citet{Mayor1987}). Furthermore, in \hb\ stars, this could be observed as a change in the morphology of the periastron variation due to a change in the inner binaries inclination. During the analysis of KIC\,3749404 we did not detect a clear signature of nodal precession in the light curve given our uncertainties.

\subsection{Stellar Components}

As predicted by theory, the gravity brightening exponent for the radiative, primary component, determined through our models, is $\sim$1 \citep{vonZeipel1924}. For the secondary component the posterior distribution is a slightly skewed normal distribution. By considering the peak of the skewed distribution (not the Gaussian fit to the posterior distribution) we find the gravity brightening exponent to be $\beta$\,=\,0.59 $\pm$\ 0.08. While this is not in agreement with the value of $\beta$\,=\,0.32, suggested by \citet{Lucy1967}, we find our value in close agreement ($\sim$2\,$\sigma$) with that of \citet{Claret2011}, who computed $\beta$\,=\,0.48 for comparable models. This value was computed using ATLAS models \citep{Castelli2004} and the \kep\ bandpass, and taking into account local gravity and convection. It is also likely that the tidal distortion affects the gravity darkening exponent, as discussed by \citet{Espinosa2012}.

In our binary models we fixed the primary and secondary rotation rates to those obtained through spectral fitting, $v\sin i_1$\,=\,$29(2)$\,$\kms$ and $v\sin i_2$\,=\,$9(2)$\,$\kms$. Combining these values with our model-determined orbital inclination (assuming that the orbital and stellar rotational axes are aligned) and radii, for the model with pulsations, we obtained $F_1$\,=\,7.1(2) and $F_2$\,=\,4.0(1), where $F$ is the ratio of the stellar rotational to orbital period and the subscript denotes the primary and secondary components, respectively. For both models, the one with, and the one without pulsations, we obtained the same values with slightly different uncertainties (see Table\,\ref{tab:ParamFree}). Interestingly, under the assumption that the stellar rotational and orbital axes are aligned, the primary component is rotating slightly faster than the predicted pseudo-synchronous rotation rate \citep{Hut1981}, the stellar rotation synchronous with the orbital velocity at periastron, which was calculated to be $F$\,=\,5.3(2) for both components; while the secondary component is rotating slightly slower. However, if we relax the assumption that the orbital axes are aligned, it is possible to find configurations where both stars are rotating pseudosynchronously. 

\section{Summary and Conclusions}
\label{sec:Conc}

We have modelled the \hb\ star binary system KIC\,3749404 by combining a comprehensive assortment of tools including the binary modelling software, \ph; \emcee, a {\sc python} implementation of the affine invariant Markov chain Montie Carlo techniques; and our own software to fit pulsations and Doppler boosting. With these tools we were able to obtain two successful fits to the \hb\ star light and radial velocity curves simultaneously: one without pulsations and one with all the tidally induced pulsations greater than 20\,$\mu$mag (7 pulsations). From these fits we obtained the fundamental parameters of the binary system (see Table\,\ref{tab:ParamFix}). The difference between the results of these two models hinges on the different inclination values: 60.2(1)\degs\ without pulsations and 62(1)\degs\ with pulsations. This 2 sigma difference subsequently produced a larger semi-major axis and masses in the model with pulsations (3(1)\,$\Rsun$ larger for the semi-major axis, and 0.3(1)\,$\Msun$ and 0.1(1)\,$\Msun$ larger for the primary and secondary component, respectively). 

Analysis of the stellar spectra highlighted a mismatch between the spectral templates and observations for the primary component. Increasing the metallicity improved the general fit to the metal lines; however, worsened the fit to the Ca II K lines and the hydrogen lines, which are important temperature indicators in A stars. This selective enrichment of the photosphere matches the abundance patterns of Am stars. The presence of Am stars in binaries is common, due to the slow rotation (compared to single A stars) and increased convective mixing. Thus we conclude that the primary component of KIC\,3749404 is an Am star.

As there is significant smearing of the periastron variation in the phased \kep\ light curve, which is indicative of apsidal motion, we elected to model the rate of apsidal advance. This was done by fitting two new models, one to the beginning (Quarters 1 and 2) and one to the end (Quarters 15 and 16) of our data set. For the new models, all the values were fixed to those previously determined for the model with pulsations, except for the argument of periastron and phase shift, which were fitted, along with an additional parameter to shift all the pulsation phases by a fixed amount. The difference between the argument of periastron at the beginning and end of our data set, divided by the duration of the data set gave us an estimate of the rate of apsidal advance ($\dot{\omega}$$_{obs}$\,=\,1.166(1)\degs/yr). Comparing this value to the theoretical rate of apsidal advance, accounting for both classical and general relativistic effects ($\dot{\omega}$$_{theo}$\,=\,0.007(6)\degs/yr), we found that the orbit of KIC\,3749404 is precessing faster than predicted by two orders of magnitude.  While we accept that the lack of eclipses in our light curve limits the determination of the stellar radii and rate of apsidal advance, the extreme disagreement between theory and observation is unlikely a consequence of our chosen methods, even when considering that the rate of classical apsidal advance scales as $R^{-5}$. After eliminating tidally induced pulsations as the sole cause of rapid apsidal motion, we hypothesise that it is due to the presence of a tertiary component in the system. 

\section{Acknowledgements}
The authors express their sincere thanks to NASA and the \kep\ team for allowing them to work with and analyse the \kep\ data making this work possible. The \kep\ mission is funded by NASA's Science Mission Directorate. KH and AP acknowledge support from the NSF grant \#1517460. This work was also supported by the STFC (Science and Technology Funding Council). The authors would like to thank Ed Guinan for the enlightening discussion on apsidal motion. KH, ST and JF acknowledge support through NASA K2 GO grant (11-KEPLER11-0056). We would like to thank the RAS for providing grants which enabled KH's attendance to conferences and thus enabled the development of collaborations and the successful completion of this work. AP acknowledges support through NASA K2 GO grant (NASA 14-K2GO1\_2-0057). For the spectroscopic observations and results from TRES, D.W.L. and S.N.Q acknowledge partial support from the Kepler mission under NASA Cooperative Agreement NNX11AB99A to the Smithsonian Astrophysical Observatory. This work was performed in part under contract with the Jet Propulsion Laboratory (JPL) funded by NASA through the Sagan Fellowship Program executed by the NASA Exoplanet Science Institute.

%     R E F E R E N C E S

\bibliographystyle{mn2e}
\bibliography{StarRef_2} 
\label{lastpage}

\end{document}